# PHYSICAL SIMULATION OF NUCLEATION AND CRYSTALLIZATION PROCESSES IN TRANSPARENT ORGANIC MELTS

A.S Nuradinov[1], A.V. Nogovitsyn[1], K.A. Sirenko[1], I.A. Nuradinov[1], O.V. Chystiakov [1], D.O. Derecha[1,2]*

[1] *Physico-technological Institute of Metals and Alloys Natl. Acad. Of Sci. of Ukraine. Acad. Vernadskoho 34/1, Kyiv, Ukraine, 03680.*
[2] *Institute of Magnetism Natl. Acad. Of Sci. of Ukraine and Min. of Edu. and Sci. of Ukraine. Acad. Vernadskoho 34, Kyiv, Ukraine, 03142.*

*Corresponding author: Dmytro Derecha. E-mail: dderecha@gmail.com

**Abstract**

The aim of this work is the development of scientifically based ways of influencing the process of nucleation of crystallization centres to control the cast structure and the properties of blanks at the casting stage. The possibility of controlling of crystals nucleation was studied on transparent modelling media using the droplet method. The dependence of the crystallization kinetics of drops with a size of 100÷200 μm on the overheat temperature, the exposure in the overheated state and the degree of supercooling was established. The curves of the crystallization kinetics of drops from their melts as the criteria for evaluating the crystallization process were chosen.

The nature of the change in the experimental media crystallization kinetics curves confirms the assumption about the decisive role of impurities in the crystallization processes of any alloys. The confirmation of the hypothesis about the decisive role of impurities in the processes of crystallization from positions of heterogeneous nucleation is the dependence of the number of solidified drops with the same diameter on the degree and duration of their overheating.

**Introduction**

In the production of metallurgical blanks, the crystallization of relatively large volumes of metal occurs. The melting and crystallization regimes are the key technological parameters that affect the quality of the cast product. The degree of supercooling of melt relative to its melting temperature before the crystallization has a decisive influence on the quality of the cast metal. The effect of supercooling on the quality of the blank is due to the following the more supercooling is required to start crystallization the smaller the grain size of the metal [1, 2].

In this case, the reduction of the grain size leads to the mechanical characteristics growth of the cast metal [3-5]. It is known that, in the common case, with the increase of overheating the supercooling increases with the attainment of saturation [6-10].

There are various theories explaining the dependence of melt supercooling on overheating. They can be conditionally divided into theories of surface and volume crystallization. In the first case, it is considered that crystallization begins on the surface between the melt and the crucible (or oxide film), and overheating affects the condition of this surface. It is assumed that solid metal remains in the pores of the crucible or the pores of the oxide film up to a certain critical temperature of overheating due to the surface energy and the increased pressure that it experiences in the pore due to the higher coefficient of expansion of the metal than that of the oxide [6, 7]. This theory can explain the change in the mechanism of crystallization at insufficient overheating: up to a few units or a maximum of tens of degrees above the melting temperature. But this explanation is not suitable, for example, for aluminium, for which maximum supercooling is achieved at overheating of more than 200 °C [11, 12].

An alternative theory asserts the existence of micro-groups of atoms (clusters) in the melt, which are preserved when the melt is slightly overheated above the liquidus and serve as nuclei during crystallization but are destroyed when the melt is overheated. But the cluster theory was also not experimentally confirmed for pure aluminium [3-5, 7].

The crystal nucleation mechanisms investigation in real objects is practically impossible due to the high temperature and opacity of metal melts, the thermal and hydrodynamic processes controlling impossibility, etc. These circumstances caused the need to conduct the investigations using indirect methods such as physical simulation on transparent organic media [11-14, 20].

To put the processes control practical recommendations on forming the structure and properties of cast blanks on a scientific basis, reliable methods for determining the number of crystallization centres are necessary. In our opinion, in this regard, the most correct is the method of microvolumes, which consists of dividing a massive sample of melt into a large number of small drops. Therefore, in this work, the dependence of the crystallization kinetics of the drops on the degree of their overheating and supercooling using the droplet method (drops size $\geq 50$ μm) on transparent organic media was investigated.

**Materials and Techniques**

For the physical simulation, the model media was chosen based on the following considerations [13, 14]: transparency – to ensure the visualization of the investigated processes; low crystallization speed – for the accuracy of the measurements; the melting temperature below 100°C – for the experiment convenience; non-toxic - to ensure safety, etc.

Taking into account indicated requirements there are organic substances such as paraffin, camphene, salol, and diphenyl qua a medium for nucleation physical simulation processes that were used (Table 1).

Table 1. Physical properties of model medium [13, 14]

| Material | Liquidus temperature, $t_l$, (°C) | Crystallization range, $\Delta t_{cr}$, (°C) | Crystallization heat, $Q_o$, (KDj/кп) | Calorific conduction, $\lambda$, W/m,·(°C) | Specific heat, C, (KDj/kg·°C) | Dencity, $\rho$, (kg/m³) | Thermal diffusivity, $a$, * m²/sec |
|---|---|---|---|---|---|---|---|
| Paraffin | 52 | 0 | 150,1 | 0,24/0,19 | 3,22/ 3,86 | 812/764 | $9,3 \cdot 10^{-8}$/ $7,5 \cdot 10^{-8}$ |
| Camphene | 45 | 3 | 138 | - / 1,85 | - / 2,9 | 845/815 | - / $14,2 \cdot 10^{-8}$ |
| Salol | 43 | 0 | 91 | -/4,016 | -/0,18 | 1250/1195 | $7,2 \cdot 10^{-7}$/ $7,5 \cdot 10^{-7}$ |
| Diphenyl | 67 | 6 | 144,4 | 0,42/0,53 | 2,08/2,66 | 1135/1035 | $17,9 \cdot 10^{-8}$/ $15,5 \cdot 10^{-8}$ |

*Note: value in solid state / value in liquid state.*

When conducting research it was necessary to obtain the studied medium in the form of dispersed drops on a flat surface. For this drops of experimental medium melt were sprayed by an air atomizer onto the glass substrate with the dimensions of 20x20 mm which was placed under a special cap. The diameter of the drops under crystallization was in the range of 100÷200 μm.

The process of crystallization of drops was studied on the installation presented in Fig. 1. A glass substrate with dispersed drops of the experimental medium melt (1) was placed in a rectangular cell (2) connected to a thermostat. A cell was isolated from contact with an atmosphere by plug (3). The required temperature inside the cell was maintained by a thermostat with a thermocouple (4) installed into the plug. A cell was placed on the stage of a microscope (5). The observation was conducted in a transmission configuration (6). The images were recorded by a video camera (7) installed on the microscope.

At regular intervals of time ($\tau$) the number of solidified drops ($n_s$) was determined, and their relative number ($\eta = \frac{n_s}{n_0} \times 100\%$) was calculated.

According to diameter, all drops were divided into three groups: the first group – 100±20μm; the second – 150±20μm; the third – 200±20μm. Up to 10 drops ($n_o$) were observed in each group.

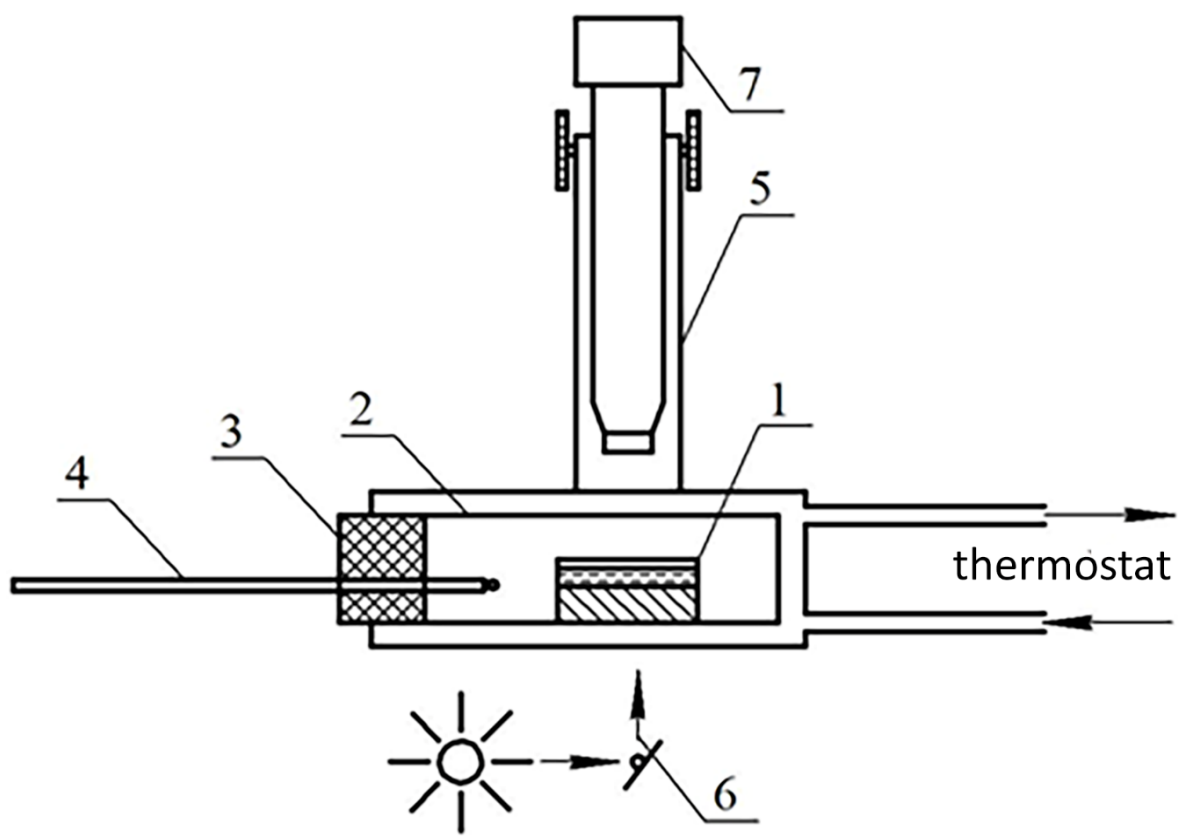

Figure 1. Scheme of experimental setup.

1 – substrate; 2 – cell; 3 – plug; 4 – thermocouple; 5 –microscope stage; 6 – microscope lamp; 7 – video camera.

## Results and Discussion

Considering that the temperature of the crystallization process of the drops beginning was evaluated by the air temperature in the middle of the cell, at the first stage the time of establishing the temperature regime inside the cell was determined.

Since the atmosphere inside the cell was air, there was always (both during heating and cooling) a difference between the temperatures of the coolant cooling the cuvette and the temperature inside the chamber. The maximum time to reach the required supercooling of the studied media (drops of them), regardless of the depth of supersaturation, did not exceed 1 minute. This particular one made it possible to more accurately evaluate the moment when drops crystallization process starts.

Crystallization, like any mass transfer process, is characterized by statics and kinetics. Statics determines the equilibrium between the melt and the crystals that form. Kinetics determines the speed of the crystallization process, i.e., shows the resulting crystals' linear growth quantity and quality (size, fractional composition, etc.). Figure 2 shows images of drops of paraffin and salol melts before the moment of supercooling (supersaturation) and after regular time intervals after the start of their crystallization. As we can see until the moment of supersaturation (that is, reaching the crystallization temperature), all drops are in a liquid state, regardless of their dispersion (Fig. 2). At the same time, all paraffin drops solidified within 300 seconds with insignificant supercooling (3°C). (Fig. 2c), but for salol, a certain number of drops for the same moment (for 300 seconds) is in a liquid state (Fig. 2f), although the degree of their supersaturation was much higher (10°C). All this indicates is the different purity of the test media and clearly shows the decisive role of impurities in the crystallization processes. Indeed, in these experiments, the purity of paraffin is technical ($\leq$ 95%) and the salol is analytically pure. (~ 99%).

The criteria for the evaluation crystallization process of the media under investigation were the curves of the crystallization kinetics of drops from their melts. Figure 3 shows the curves of the paraffin and salol melt drops crystallization kinetics overheated at 5°C for 1 min. It can be seen that the nature of the change in $\eta$ depending on the cooling time $\tau$ is the same for both environments. Other things being equal, the value of $\eta$ for drops of a larger diameter is significantly higher, and for extreme drop sizes, this difference reaches the level of 40÷60% (Fig. 3, curves 1 and 3, 4 and 5). Such a dependence of the parameter η on the size of the drops, in our opinion, confirms the assumption about the decisive role of impurities in the crystallization processes of any alloys. The logical consequence of this phenomenon should be the dependence of the number of solidified drops of experimental media of the same diameter on the overheating degree and time.

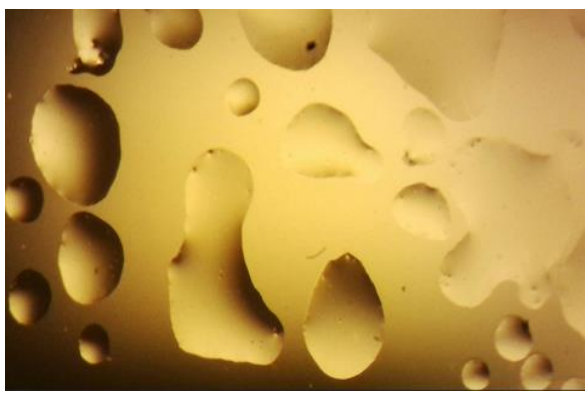

a)

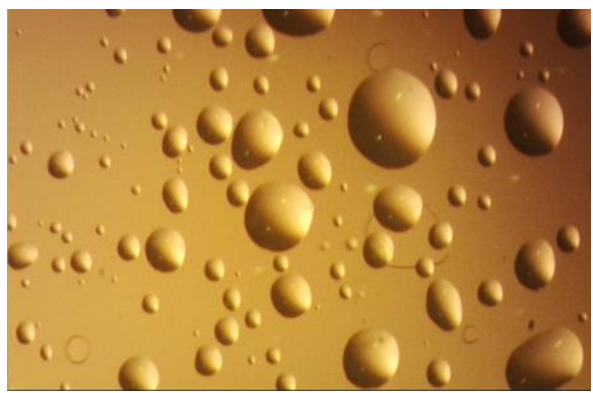

d)

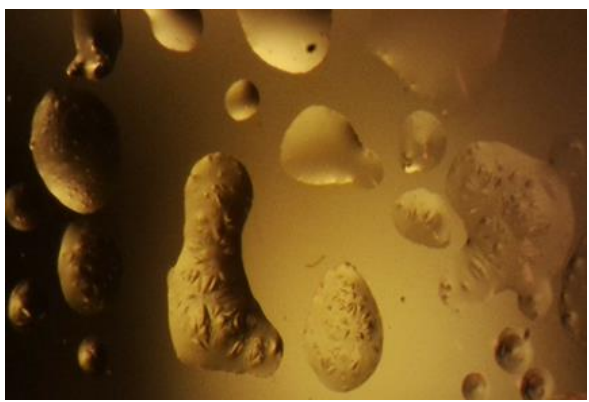

b)

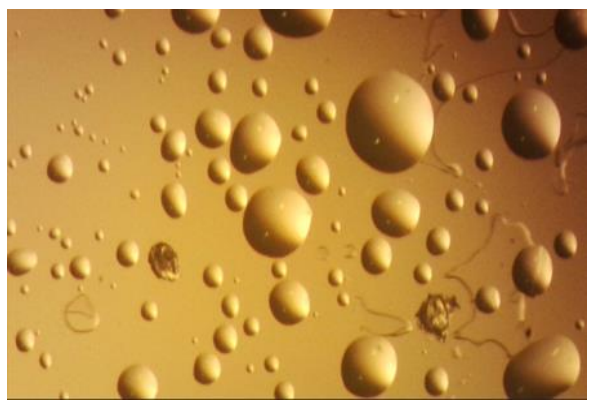

e)

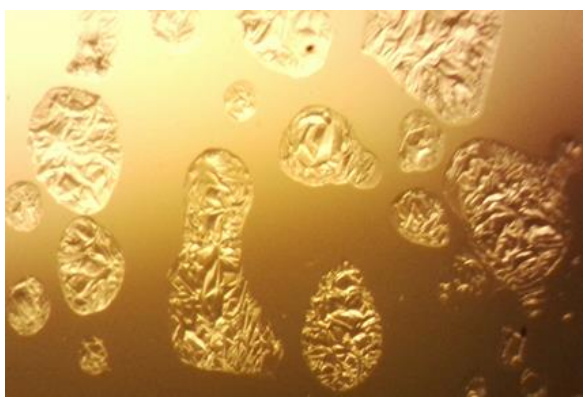

c)

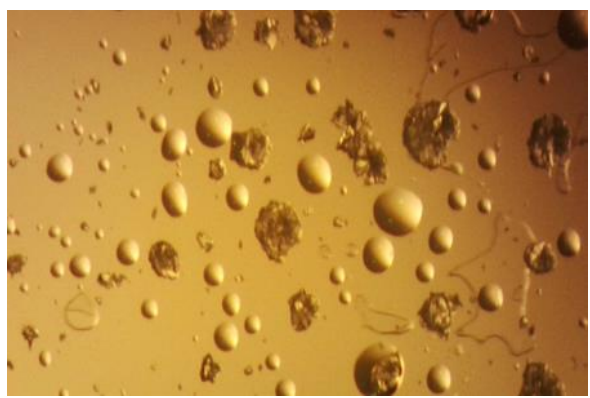

f)

Figure 2. Crystallization of paraffin drops (a, b and c) and salol (d, e and f) during supercooling at 3°C and 10°C respectively. Time intervals: a, d – 0 sec.; b, e – 120 sec.; c, g – 300 sec.

Experiments showed that the number of drops that crystallized before the temperature in the chamber was set to $\leq t_m$ (melting temperature) significantly depends on their size, temperature of overheating, time of exposure in the overheated state, and degree of supersaturation (Fig. 4, start points). The high probability of the presence of active impurities in melts of a larger volume explains the dependence of supercooling on the drops size. The dependence of the parameter η for paraffin drops of the same size on the overheating value and the exposure time can be explained from the same positions of heterogeneous nucleation (Fig. 4). It is likely that with greater overheating of the melt (or longer overheating), partial deactivation (dissolution) of the impurities occurs. The number of solidifying droplets changes at the same supersaturation (supercooling) value. Therefore, drops of the same size for τ = const have different values of the parameter η, even though for the crystallization of drops of such sizes it is enough to have at least one active nucleating center (Fig. 4).

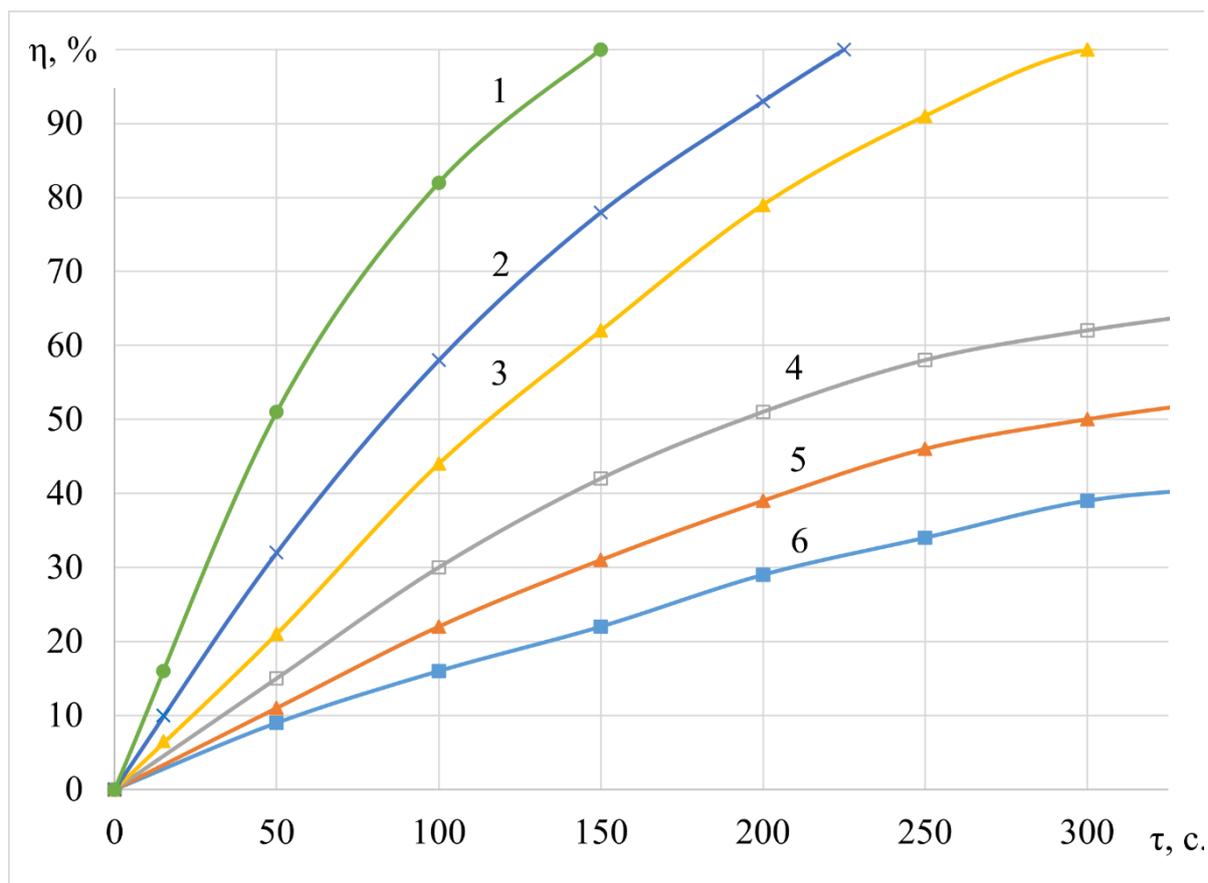

Figure 3. Kinetics of paraffin (1, 2 and 3) and salol (4, 5 and 6) drops crystallization with diameters of 100µm (3, 6), 150µm (2, 5) and 200µm (1, 4) with supercooling temperature of 3°C and 10°C, respectively.

With an increase in the superheat temperature of the paraffin melt from 5°C to 15°C, the degree of crystallization η significantly decreases (up to 40%, Fig. 4, curves 1 and 3). In addition, the growth of the exposition time of the drops in an overheated state from 1 min (dashed lines) up to 20 min (solid lines) this parameter also noticeably decreases. Such dependencies are difficult to explain otherwise than by the activity changing of impurities in the drops.

The literature is free of analytical solution that allows us to proceed from the crystallization kinetics curves of alloys to the crystallization rate. This is because it is impossible to determine the induction period of crystallization for large volumes of solidifying melts. It is an important characteristic of the crystallization process, which allows to determining the specific energy at the melt-crystal interface, the maximum supercooling of the melt, etc. Investigation of crystallization processes by the droplet method solves this problem because many identical drops can be obtained from one large volume of melt of the same composition and purity. Dispersed drops with the same diameter from the same melt are a large number of experimental samples under perfectly identical crystallization conditions.

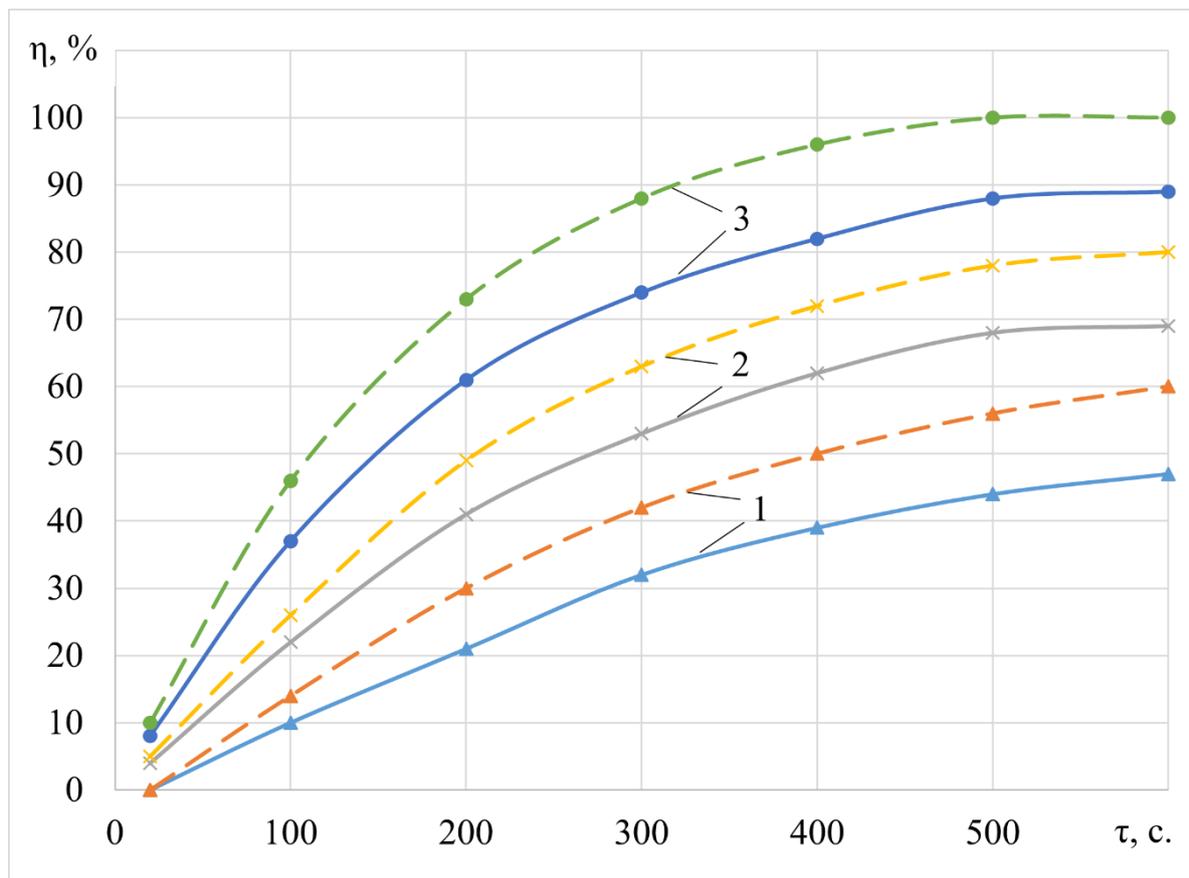

Figure 4. Crystallization kinetics of paraffin drops with a diameter of 100μm depending on the overheating temperature and the exposition time in the superheated state at 2°C and supercooling: 1 – 15°C superheating; 2 – 10°C and 3 – 5°C.

With large subcooling of melts in small volumes (as opposed to large volumes), slow nucleation and growth of separate crystals are not observed. In transparent drops of model media, it is enough for one centre of crystallization to appear and it instantly becomes cloudy (solid), that is, the induction period has a minimum duration. At the same time, the induction period of crystallization is defined as the average time required for the clouding (solidification) of all cooled drops with the same size ($\tau_s$).

The drops crystallization rate as a function of time $n_s(\tau)$ can be written as:

$$n_s(\tau) = \frac{dn_s}{d\tau},\qquad(1)$$

where $dn_s$ is the number of solidifying drops during the time $d\tau$.

The function $n_s(\tau)$ can be revealed for homogeneous nucleation since this process is random and crystallization of the one drop per unit of time ($\omega$) probability for all drops of the same size and equal supercooling is a constant value. This assumption is acceptable since all the drops observed during the experiment are in the same conditions. Then, with an unchanged value of $\omega$, the decrease in the number of unsolidified drops dn during the time d$\tau$ can be expressed by the following ratio:

$$dn = -\omega \cdot n \cdot d\tau,\qquad(2)$$

where $n$ is the number of unsolidified drops. The sign minus indicates a decrease in unsolidified drops over time.

Equation (2) can be rewritten as $dn/n = -\omega \cdot d\tau$ and integrated in the interval from $\tau = 0$ to $\tau = \infty$:

$$ln \frac{n}{n_0} = -\omega \cdot \tau + C.\qquad(3)$$

The constant of integration $C$ can be found from the condition that at the beginning of crystallization $\tau = 0$ and $n = n_0$. Then from Eq. (3) we obtain $ln\, n_0 = C$. Considering this, Eq. (3) can be presented as follows:

$$ln \frac{n}{n_0} = -\omega \cdot \tau \text{ або } n = n_0 \cdot e^{-\omega\tau}.\qquad(4)$$

Here $n$ is the number of drops that have not yet solidified at time $\tau$; $n$ is the total number of drops. The number of solidified drops $n_s$ at time $\tau$ is equal to

$$n_s = n_0 - n.\qquad(5)$$

Then, taking into account Eq. (4), equation Eq. (5) can be rewritten as

$$n_s = n_0 \cdot (1 - e^{-\omega\tau}).\qquad(6)$$

Thus, as a result of this analysis, we come to the Eq. (6) which describes the kinetic curves of droplet crystallization.

After taking the logarithm of Eq. (6), we reduce it to the following form:

$$ln(1 - \frac{n_3}{n_0}) = -\omega \cdot \tau. \qquad (7)$$

Next, will try to move from the kinetic curves of the drops crystallization to the crystallization rate I (that is, the number of crystals formed per unit of time in a unit of volume). It was noted above that the drops solidified upon the appearance of one crystallization centre. From this, assume a correspondence between the number of solidified drops n with the number of crystallization centres at the current time $\tau$. Then the rate of crystallization I can be simplified as follows:

$$I = K_D \frac{dn}{d\tau} \times \frac{1}{V}, \qquad (8)$$

where $K_D$ is the proportionality factor that takes into account the differences in the number of nuclei in the drop and the volume of the melt separated into drops; $V$ is the volume of drops that have not solidified at this point in time.

The volume of unsolidified drops at the moment of time $\tau$ is equal to

$$V = \frac{1}{6} \cdot \pi \cdot D^3 \cdot n, \qquad (9)$$

where $D$ is the drops diameter.

Taking into account Eqs. (6) and (8), expression (7) will have the following form:

$$I = \frac{6K_D}{\tau_c \pi D^3}, \qquad (10)$$

where $\tau_c = \frac{1}{\omega}$ – the drops average lifetime in the liquid state.

According to the molecular kinetic theory, the rate of heterogeneous nucleation of crystals in a certain volume of supercooled melt I is described by the equation [17-19]:

$$I = K_I exp[-\frac{U}{Rt} - \frac{\gamma \sigma^3 V_0^2 t_0^2}{kq^2 t(\Delta t)^2}], \qquad (11)$$

where $K_I$ is a pre-exponential factor; $U$ is the activation energy; $\gamma$ is the shape coefficient of crystals; $\sigma$ is the specific surface energy of the crystal–melt interface; $V_0$ is the volume of one molecule in the solid phase; $t_0$ is the melting temperature; *t* is the cooling temperature; *R* is the absolute gas constant; *K* is the Boltzmann constant; *q* is the melting heat; $\Delta t = t0 - t$ is the value of the melt supercooling.

Assuming that the entire volume of the melt and its drops are thermostated at the same temperatures, we equate the right-hand sides of Eqs. (10) and (11), logarithmize this equality and obtain the expression

$$ln\tau_c - \frac{U}{Rt} = ln\frac{6K_D}{K_I \pi D^3} + \frac{\gamma \sigma^3 V_0^2 t_0^2}{kq^2 t (\Delta t)^2}. \tag{12}$$

Thus, the analytical analysis of the obtained results made it possible to derive important equations that establish the relationship between the crystallization parameters of the drops and the temperature and time parameters of their solidification (τ, *t* and Δ*t*).

Verification of equations (7) and (12) for paraffin in a wide range of changes of η value is presented in Fig. 5 and 6. They have a linear character that qualitatively agrees with the heterogeneous nucleation theory [13, 16-19].

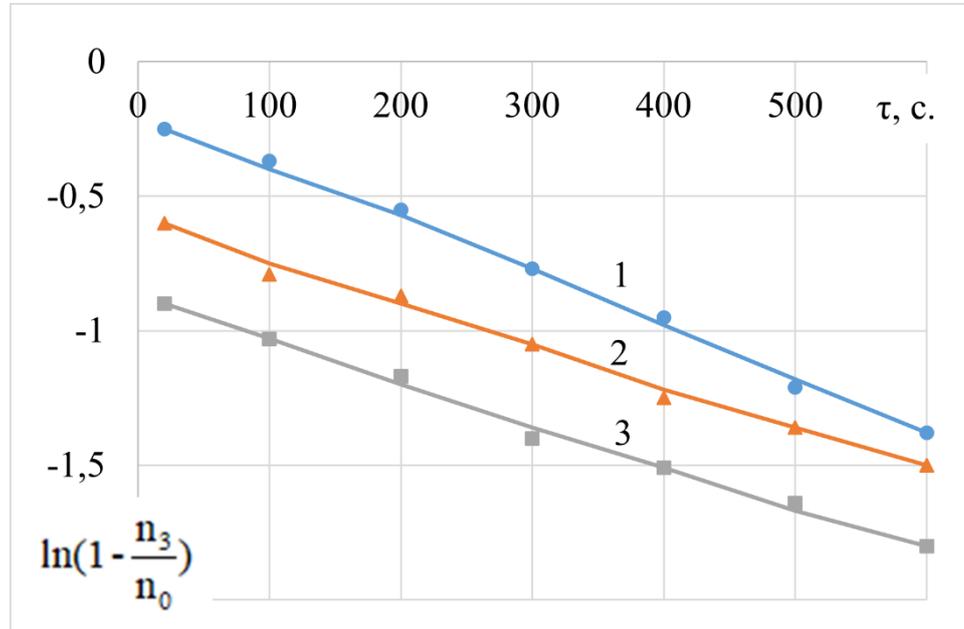

Figure 5. The change of the parameter $ln(1 - \frac{n_3}{n_0})$ on τ (Eq. (7)) for paraffin drops with a diameter of 100μm supercooled by 2°C, overheated by: 1 – 15°C, 2 – 10°C and 3 – 5°C

Using expression (12) and Fig. 6, it is possible to estimate the interphase energy at the crystal-melt boundary $\sigma^*$. The ratio of $K_D/K_I$ coefficients can be estimated by the segment cut off by a straight line on the ordinate axis (Fig. 6), and its slope to the abscissa axis allows us to find $\sigma^*$. However, it should be taken into account that the value of the interphase energy determined in this way has no physical meaning but is only a quantitative characteristic of the catalytic activity of impurities in the process of crystal nucleation [13]. This was confirmed by a comparative analysis of the obtained interphase energy values for experimental media with known literature data. This analysis showed that the interphase energy levels calculated by us using the droplet method turned out to be close to the values known in the literature for large volumes and determined in different ways (dispersive analysis of crystalline sediment in melts, determination of the stability of large volumes of melts, etc.), where heterogeneous crystal nucleation is more likely. For certain experimental environments, the calculated values of σ* turned out to be even lower than for large volumes. Accordingly, this indicates a heterogeneous mechanism of crystal nucleation in drops from the studied media in the considered range of their sizes (100÷200μm).

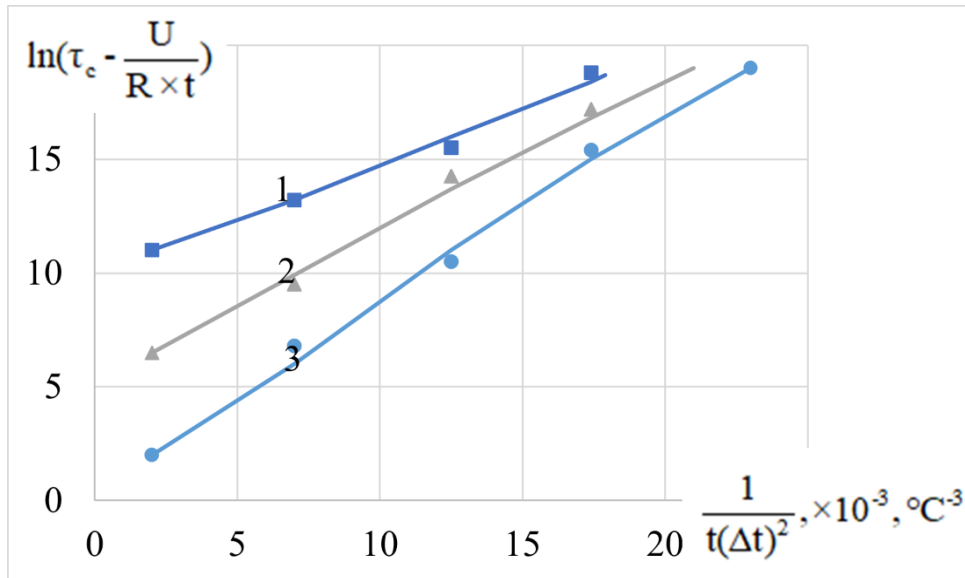

Figure 6. The change of the parameter ($\boldsymbol{ln\tau_c - \frac{U}{Rt}}$) on $\frac{1}{t(\Delta t)^2}$ (Eq. (12)) for paraffin drops with a diameter of 100 μm, superheated by 5°C, with supercooling of 2°C

**Conclusions**

Thus, the developed method of the crystals nucleation in model environments investigation by the droplet method allows to establish the dependence of the crystallization kinetics of drops with different dispersion on the degree of their overheating and supersaturation (supercooling). The analysis of the obtained results indicates a heterogeneous mechanism of crystal nucleation even in drops of microscopic size (~100 μm). The criteria for evaluating the process of solidification of the investigated media were the curves of the kinetics of crystallization of drops, the nature of which changes confirm the assumption about the decisive role of impurities in the crystallization processes of any alloys. The analysis of these curves indicates a heterogeneous mechanism of crystal nucleation even in droplets of microscopic size (~100 μm). Otherwise, it is difficult to explain the fact of a decrease in the number of drops of the same size that harden, with an increase in the degree of their overheating or the time of exposure in an overheated state, other things being equal. And here, in our opinion, it is not a matter of mechanical impurities (solid substrates), which under certain conditions themselves become centres of crystallization, but of limitedly soluble impurities. The degree of solubility of such impurities is precisely determined by the temperature of overheating and the time the melt is kept in this state. This technique for studying the processes of nucleation and growth of crystals depending on their mass (size) in model alloys can be applied in real conditions of mass crystallization.

**References**

1. Qian M., Cao P., Easton M.A., McDonald S.D., StJohn D.H. An analytical model for constitutional supercooling-driven grain formation and grain size prediction // Acta Materialia. – 2010. – V. 58, No. 9. – P. 3262–3270. (DOI: 10.1016/j.actamat.2010.01.052).

2. StJohn D.H., Prasad A., Easton M.A., Qian M. The Contribution of Constitutional Supercooling to Nucleation and Grain Formation // Metall. Mater. Trans. A. – 2015. – V. 46. – P. 4868-4885. (DOI:10.1007/s11661-015-2960-y).

3. Jeong G., Park J., Nam S., Shin S.-E., Shin J., Bae D., Choi H. The Effect of Grain Size on the Mechanical Properties of Aluminum // Archives of Metallurgy and Materials. – 2015. – V. 60, No. 2. – P. 1287-1291. (DOI:10.1515/amm-2015-0115).

4. Schempp P., Cross C.E., Häcker R., Pittner A., Rethmeier M. Influence of grain size on mechanical properties of aluminium GTA weld metal // Weld World. – 2013. – V. 57. – P. 293–304. (DOI: 10.1007/s40194-013-0026-6).

5. El-Danaf E.A., Soliman M.S., Almajid A.A., El-Rayes M.M. Enhancement of mechanical properties and grain size refinement of commercial purity aluminum 1050 processed by ECAP // Materials Science and Engineering A. – 2007. – V. 458. – P. 226–234. (DOI: 10.1016/j.msea.2006.12.077).

6. Yang B., Perepezko J.H., Schmelzer J.W.P., Gao Y., Schick C. Dependence of crystal nucleation on prior liquid overheating by differential fast scanning calorimeter // The Journal of Chemical Physics. – 2014. – V. 140, 104513. – P. 104513-1 – 104513-7. (DOI: 10.1063/1.4868002 ).

7. Zhao B., Li L., Lu F., Zhai Q., Yang B., Schick C., Gao Y. Phase transitions and nucleation mechanisms in metals studied by nanocalorimetry: a review // Thermochimica Acta. – 2015. – V. 603. – P. 2-23. (DOI: 10.1016/j.tca.2014.09.005).

8. Koh H.J., Rudolph P., Schäfer N., Umetsu K., Fukuda T. The effect of various thermal treatments on supercooling of PbTe melts // Materials Science and

Engineering B. – 1995. – V. 34B. – P. 199-203. (DOI: 10.1016/0921-5107(95)01319-9).

9. Tong H.Y., Shi F.G. Abrupt discontinuous relationships between supercooling and melt overheating // Applied Physics Letters. – 1997. – V. 70, No. 7. – P 841-843. (DOI: 10.1063/1.118220).

10. Tong H.Y., Shi F.G. Dependence of supercooling of a liquid on its overheating // The Journal of Chemical Physics. – 1997. – V. 107, No. 19. – P. 7964-7966. (DOI: 10.1063/1.475057).

11. Jinku Y., Qi Q., Lian L., Qihua J., Dongying N. Investigation on the undercooling and crystallization of pure aluminum melt by DSC // Phase Transitions. – 2010. – Vol. 83, No. 7. – P. 543–549. (DOI: 10.1080/01411594.2010.499711).

12. Mei Q., Li J. Dependence of Liquid Supercooling on Liquid Overheating Levels of Al Small Particles // Materials. – 2016. – V. 9, No. 1, article 7. – P. 1-8. (DOI: 10.3390/ma9010007).

13. Yazdanpanah, Nima and Nagy, Zoltan, eds. (2020) *Nucleation and crystal growth in continuous crystallization.* In: The Handbook of Continuous Crystallization. Royal Society of Chemistry,609p. (ISBN: 978-1-83916-131-5). (DOI: 10.1039/9781788013581).

14. Nuradinov A.S. Nahaev M.R. Prozessi kristallizazii I formirovaniya strukturi lityh zagotovok. –Grozniy: FGBOU VO ChGU. – 2020. – 170p. [in Russian].

15. Josef Kuneš Similarity and Modeling in Science and Engineering. Cambridge International Science Publishing Cambridge, 442p. (DOI: 10.1007/978-1-907343-78-0). (ISBN: 978-190-734-377-3).

R.I.L Guthrie, R.P Tavares, Mathematical and physical modelling of steel flow and solidification in twin-roll/horizontal belt thin-strip casting machines, Applied Mathematical Modelling, Volume 22, Issue 11, 1998,

16. Pages 851-872, ISSN 0307-904X, https://doi.org/10.1016/S0307-904X(98)10027-6.

17. Flemings, M.C. (1974) Solidification Processing. McGraw-Hill Book Company, New York, 423p.

18. J. Frenkel Kinetic Theory of Liquids. Peter Smith Publisher, Incorporated, 1984. (ISBN: 084-462-094-7).

19. Chalmers B. Principles of Solidification. Wiley Series on the Science and Technology of Materials, 1964 319p. (ISBN: 047-114-325-1).

20. Derecha D.O., Skirta Yu.B., Gerasimchuk I.V., Hruzevych A.V., Statistical and Fourier analysis of the vortex dynamics of fluids in an external magnetic field,Journal of Electroanalytical Chemistry, Volume 873, 2020, 114399, (DOI: 10.1016/j.jelechem.2020.114399).